# The Two-Jet Differential Cross Section at $\mathcal{O}(\alpha_s^3)$ in Hadron Collisions

W. T. Giele

*Fermi National Accelerator Laboratory, P. O. Box 500,
Batavia, IL 60510, U.S.A.*

E. W. N. Glover

*Physics Department, University of Durham,
Durham DH1 3LE, England*

and

David A. Kosower

*Service de Physique Théorique, Centre d'Etudes de Saclay,
F-91191 Gif-sur-Yvette cedex, France*

March 1994

**Abstract**

We study the two-jet inclusive cross section via the triply differential distribution $d^3\sigma/dE_T d\eta_1 d\eta_2$ to next-to-leading order in QCD. The predicted distributions can be compared directly with forthcoming data from the D0 and CDF experiments at Fermilab. We discuss differences with the leading-order predictions, and examine uncertainties due to the choice of scale and parton density.

In hadronic collisions, the most basic form of the strong interaction at short distances is the scattering of a colored parton off another colored parton. Experimentally, such scattering can be observed via its production of two jets or sprays of hadrons with large transverse energy. We therefore expect the study of two-jet events of this type to reveal the properties of the short-distance hard scattering. One method is the examination of the angular distribution of the two jets in their center of mass frame [1, 2] which provides information on the spin of the partons. Alternatively, we can probe the parton distributions by studying the rapidities of the two jets in the laboratory frame. In particular, by fixing the rapidity and transverse energy of one of the jets and varying the other jet rapidity, we can probe the whole range of parton momentum fractions, and possibly constrain the large-$x$ parton distributions.

In this Letter, we report on a calculation of the two-jet inclusive cross section $d^3\sigma/dE_T d\eta_1 d\eta_2$ to next-to-leading order in the strong coupling constant, that is to $\mathcal{O}(\alpha_s^3)$. Compared to that in a leading-order [$\mathcal{O}(\alpha_s^2)$] calculation, the theoretical uncertainty is reduced in the following ways:

- Greater sensitivity to the jet algorithm;
- Reduced dependence on the renormalization and factorization scales;
- Removal of kinematic constraints.

The first improvement is a consequence of admitting configurations with three partons in the final state — a jet may now be formed by the merging of two partons. This is the first step towards the recreation of an all-orders partonic jet and introduces a dependence on the size of the cone used to define the jet. The second improvement reflects the fact that these unphysical scales affect predictions of physical cross sections only because of the truncation of the perturbation series, and the next-to-leading order term pushes the truncation, and hence the unphysical sensitivity, to yet higher order: there is a partial cancellation of the scale dependence between the $\mathcal{O}(\alpha_s^2)$ and $\mathcal{O}(\alpha_s^3)$ contributions to the cross section. This, too, is a first step towards a scale-independent all-orders result. Thirdly, by admitting radiation into the final state, artificial kinematic constraints due to the $2 \to 2$ nature of the leading-order prediction are relaxed.

The results presented here represent the first application of a very general $\mathcal{O}(\alpha_s^3)$ Monte Carlo program for one, two and three jet production based on the one-loop $2 \to 2$ and the tree level $2 \to 3$ parton scattering amplitudes [3, 4]. In order to cancel the infrared singularities, the divergent regions where two partons are collinear or a gluon is soft are removed analytically from the $2 \to 3$ parton cross section using the techniques described in refs. [5, 6]. (For other techniques see ref. [7].) These divergences are precisely matched by singularities in the one-loop $2 \to 2$ matrix elements, and may be cancelled algebraically. The resulting finite $2 \to 2$ and $2 \to 3$ parton processes are then evaluated numerically and passed through a jet algorithm to determine the one and two jet cross sections according to the experimental cuts. Different cuts and/or jet algorithms can easily be applied to the parton four-momenta and to any infrared-safe distribution computed at $\mathcal{O}(\alpha_s^3)$. Furthermore, a detector response function can be applied to the raw jet momenta to obtain the observed jet momenta.

Previous calculations have focussed on the next-to-leading order corrections to the single jet inclusive transverse energy distribution [8, 9] and to the two-jet inclusive invariant mass distribution [1]. We have checked that our program reproduces the $\mathcal{O}(\alpha_s^3)$ one-jet inclusive cross section of ref. [8, 10], which agrees well with the data from CDF [11]. We have also checked that our results are independent of the unphysical parameter used to isolate the divergences [5, 6].

We wish to consider the process,

$$p\bar{p} \to \text{jet}_1 + \text{jet}_2 + X, \tag{1}$$

which can be described by the triply differential distribution $d^3\sigma/dE_T d\eta_1 d\eta_2$ where $E_T$ is the transverse energy of the leading jet, while $\eta_1$ and $\eta_2$ are the pseudo-rapidities of the jets in the laboratory frame. From these, we can determine the pseudo-rapidity of the two-jet system in the lab,

$$\eta_{\text{boost}} = \frac{1}{2}(\eta_1 + \eta_2), \tag{2}$$

and the pseudo-rapidities of the jets in the jet-jet center-of-mass frame,

$$\eta^* = \frac{1}{2}(\eta_1 - \eta_2). \tag{3}$$

At lowest order, this determines the momentum fractions of the colliding partons,

$$x_{1,2} = \frac{2E_T}{\sqrt{s}} \cosh(\eta^*) \exp(\pm\eta_{\text{boost}}). \tag{4}$$

If we now require that the leading 'trigger' jet lies in the central region $\eta_1 \sim 0$, then the rapidity of the second 'probe' jet essentially fixes the momentum fraction. In particular, when $|\eta_2|$ is large, the momentum fraction may be close to unity. For example, for hadronic collisions at $\sqrt{s} = 1800$ GeV, when $E_T = 100$ GeV and $\eta_1 = 0$, $x = 1$ for $\eta_2 = 1.42$.

A slight subtlety arises since at leading order the transverse energy of the two jets are equal. In the three parton events present at next-to-leading order, the jets no longer balance exactly. This equality is approached in events containing two hard partons while the third parton is soft. The assignment of which jet is hardest is thus dependent on the soft particles in the event and is not infrared safe. However, by interchanging the roles of the trigger and probe jets so that each event is effectively counted twice, this problem can be overcome. There is still a slight ambiguity in three jet events where the relative ordering of the second and third hardest jets with transverse momenta $E_{T2}$ and $E_{T3}$ determines which pseudo-rapidity slice ($\eta_2$ or $\eta_3$) is chosen. However, this is a higher-order effect and should be small.

To make the connection with experimental data, we consider the cross section,

$$\left\langle \frac{d\sigma}{dE_T} \right\rangle = \frac{1}{\Delta\eta_1} \int d\eta_1 \frac{1}{\Delta\eta_2} \int d\eta_2 \frac{d^3\sigma}{dE_T d\eta_1 d\eta_2}, \qquad (5)$$

where the trigger jet is fixed to lie in the central region, typically $|\eta_1| < 1$, while the probe jet may lie in different slices of $|\eta_2|$. Both the CDF [12] and D0 collaboration at Fermilab are investigating this distribution. The CDF collaboration requires the trigger jet to have a transverse energy of at least 45 GeV and to lie in the rapidity interval $0.1 < |\eta_1| < 0.7$, where the hadronic calorimeter is well calibrated and which excludes the crack in the central region. The probe jet can be studied out to a much larger rapidity, $|\eta_2| < 3.0$. The D0 collaboration also requires a trigger jet with transverse energy of at least 45 GeV, however the central trigger jet must lie in the rapidity interval $|\eta_1| < 1.0$, while the probe jet may have a rapidity up to $|\eta_2| = 4.0$. The results for the CDF measurement together with the preliminary data reported in ref. [12] are shown in fig. 1 while fig. 2 gives the predictions for the D0 experiment. We have used the standard cone algorithm [13] with a cone size of 0.7 to define the jet. For the parton distributions, we have chosen the improved MRSD$-$ set of ref. [14] which approximately reproduces the low-$x$ behaviour of $F_2^{ep}$ as measured at HERA. We used the running *one*-loop strong coupling constant $\alpha_s$ in calculating the leading-order predictions, and the two-loop running coupling constant for the next-to-leading order predictions. In both cases, we have taken $\Lambda_{QCD}^{(4)} = 230$ MeV as specified by the structure function parametrization, so that $\alpha_s^{(1)}(M_Z) = 0.131$ and $\alpha_s^{(2)}(M_Z) = 0.111$. Both the renormalization and factorization scales have been chosen to be the average $E_T$, $\langle E_T \rangle$, of jets passing the trigger jet requirements.

From figs. 1 and 2 we see that for central production of the probe jet ($|\eta_2| < 1.5$) the corrections to the leading order predictions are small for this scale choice over the whole transverse energy range of the trigger jet[1]. For larger rapidities we observe large corrections even for moderate transverse energies. If these corrections were due solely to the presence of higher-order terms in the QCD perturbative expansion this would signal a breakdown of perturbation theory at large rapidities. We will argue that this is not the case, and that this enhancement is due to a kinematic restriction imposed by the lowest order $2 \to 2$ parton scattering. Using eq. 4 we can calculate the maximum transverse energy obtainable in the leading-order cross section. The results for both the CDF and D0 cuts are listed in Table 1. At next-to-leading order, additional radiation can alter the two-jet configuration significantly by reducing the boost and thereby evading the lowest-order kinematic constraint. Nevertheless, even in three parton final states there remains a kinematic limit on the maximum transverse energy of the trigger jet which is also listed in table 1. Configurations close to this limit correspond to events where $E_{T2} \sim E_{T3} \sim \frac{1}{2} E_T^{\text{NLO max}}$ and $\eta_3 \sim -\eta_2$ and where it should be possible to identify three distinct jets. This limit can be seen in figs. 1 and 2: once the transverse energy reaches approximately half $E_T^{\text{LO max}}$ the next-to-leading order cross section starts to deviate from the leading-order cross section. Close to the kinematic limit the deviations become very large since the leading-order cross section is forced (artificially) to zero. It is therefore clear that the large corrections do not signal a problem within the perturbative expansion but are purely due to phase-space effects: two-parton final states are too restricted to describe results at large rapidities. (Large infrared logarithms would emerge from a kinematic constraint on the three parton configurations rather than on the two parton configurations.) In these regions it is necessary to include the next order in theoretical calculations. Provided we stay well below the three-parton kinematic boundary the predictions should be reliable and should agree with the data. For large rapidity differences, at the upper end of those considered here or beyond, it is probably necessary to resum logarithms in the virtual corrections [15].

---

[1] The Monte Carlo integration over phase space of course yields results with statistical errors. The curves in all the figures were obtained by fitting a smooth function through the obtained results.

| $\Delta\eta_2$ | $E_T^{LO\ MAX}$ | $E_T^{NLO\ MAX}$ |
|---|---|---|
| 0.1-0.7 (CDF) | 813 | 895 |
| 0.7-1.2 (CDF) | 578 | 816 |
| 1.2-1.6 (CDF) | 408 | 658 |
| 1.6-2.0 (CDF) | 298 | 515 |
| 2.0-3.0 (CDF) | 211 | 385 |
| 0.0-1.0 (D0) | 900 | 900 |
| 1.0-2.0 (D0) | 484 | 708 |
| 2.0-3.0 (D0) | 214 | 378 |
| 3.0-4.0 (D0) | 85 | 163 |

Table 1: The maximum allowable transverse energy, $E_T^{\max}$, of the trigger jet given the probe jet at leading (LO) and next-to-leading order (NLO) in a given rapidity bin.

For comparison the CDF preliminary data [12] was added to fig. 1. The data were corrected for detector effects and can therefore be compared directly to the theoretical predictions. Note that the systematic error is not included. As the figure shows, the inclusion of next-to-leading order corrections, lifting the kinematic constraint on the transverse energy of the trigger jet, is needed to describe the data.

To study the theoretical uncertainties, we varied the renormalization (and equal factorization) scale by a factor of two around the central value of $\langle E_T \rangle$. The integrated cross section for these scale choices are shown in table 2. The inclusion of higher-order corrections reduces the uncertainty substantially over a large region of phase space. However, as soon as the next-to-leading order distribution approaches the leading-order kinematic limit as given in table 1 the cross section is dominated by the three-parton contribution and is basically a leading-order prediction. In these regions, a strong sensitivity to the scale reappears, and a yet-higher order calculation would be required to reduce this sensitivity. Apart from this effect we see that changing scales in the range indicated varies the leading order result by $\pm 30\%$ (with some effects on the shape of distributions as well), whereas at next-to-leading order a 10% variation results with essentially no effect on the shape of the distribution.

The theoretical predictions also have a non-trivial dependence on the parton distributions.

| $\Delta\eta_2$ | $\sigma^{\rm LO}$ (nb) | $\sigma^{\rm NLO}$ (nb) |
|---|---|---|
| 0.1-0.7 (CDF) | $105.5^{+36.4}_{-25.3}$ | $99.7^{+7.1}_{-10.0}$ |
| 0.7-1.2 (CDF) | $99.2^{+35.7}_{-24.4}$ | $91.2^{+4.4}_{-8.7}$ |
| 1.2-1.6 (CDF) | $84.8^{+32.2}_{-21.6}$ | $77.6^{+4.7}_{-7.1}$ |
| 1.6-2.0 (CDF) | $63.7^{+25.7}_{-16.9}$ | $60.9^{+2.2}_{-7.4}$ |
| 2.0-3.0 (CDF) | $23.7^{+10.4}_{-6.6}$ | $23.5^{+2.9}_{-3.2}$ |
| 0.0-1.0 (D0) | $102.9^{+36.1}_{-24.9}$ | $96.3^{+6.5}_{-9.2}$ |
| 1.0-2.0 (D0) | $75.9^{+29.3}_{-19.6}$ | $69.5^{+3.8}_{-6.9}$ |
| 2.0-3.0 (D0) | $22.8^{+10.0}_{-6.4}$ | $22.6^{+3.0}_{-2.9}$ |
| 3.0-4.0 (D0) | $0.43^{+0.23}_{-0.14}$ | $0.85^{+0.49}_{-0.12}$ |

Table 2: The estimated uncertainty in the overall normalization of the leading order and next-to-leading order cross section $\int dE_T \left\langle \frac{d\sigma}{dE_T} \right\rangle$ for $E_T > 45$ GeV due to the choice of the renormalization/factorization scale. The up arrow corresponds to a scale choice of $\mu_F = \mu_R = \langle E_T \rangle / 2$, while the down arrow corresponds to $\mu_F = \mu_R = 2\langle E_T \rangle$. The statistical error is approximately 1%.

However, unlike the case of the scale dependence, we do not expect the inclusion of the next-to-leading order QCD corrections to reduce this uncertainty significantly. To indicate the dependence on the parton densities we also performed the calculation replacing the favored parametrization MRSD− by MRSD0 which has a smaller gluon contribution at small momentum fractions [14]. The difference at both leading order and next-to-leading order between MRSD− and MRSD0 depends on the rapidity slice that the probe jet occupies. At large transverse energies the differences are very small. This is as expected, since the choice of factorization scale is close or equal to the transverse energy of the trigger jet. The choice of a large scale samples the parton densities after perturbative evolution over a large range of energy scales, an evolution which essentially erases all differences between the low-energy input parametrizations of the parton densities. Furthermore, the parton momentum fractions are large at high transverse momentum, and thus the different parton distribution sets yield nearly identical predictions. This is no longer true at smaller transverse energies, and the MRSD0 parametrization gives a significantly higher cross section — by as much as 20% in the central region for $E_T \sim 50$ GeV. At both leading order and next-to-leading order the observed differences remain basically the same within statistical error.

The scale uncertainty can also be reduced by considering the cross section relative to that in the central region. In other words, consider the ratio (for CDF)

$$R = \left\langle \frac{d\sigma}{dE_T} \right\rangle \Big/ \left\langle \frac{d\sigma}{dE_T} \right\rangle_{0.1<|\eta_2|<0.7} . \qquad (6)$$

(For D0, take $|\eta_2| < 1$ in the denominator.) The scale uncertainty is essentially independent of $\eta_2$ and is therefore reduced significantly in the ratio. Indeed, the residual variation in the cross section for the scale variation considered earlier is less than 13% at lowest order, while at next-to-leading order it is less than 5%. The results are shown in figs. 3 and 4, where for comparison, we have added the preliminary CDF data to fig. 3. As expected, the next-to-leading order prediction agrees better with the data at large $E_T$ and large $\eta_2$. On the other hand, we note that the structure function dependence at small $E_T$ is not reduced significantly.

In this letter we showed that for study of two-jet cross sections where one of the jets has a large rapidity the use of next-to-leading order cross sections is essential. This indispensability arises from the artificial kinematic limitations imposed by the two-parton final state (and hence by a leading-order calculation). We have also shown that the sensitivity to the renormalization (and factorization) scale at next-to-leading order is reduced significantly, and only reemerges as we approach (or once we exceed) the two-parton kinematic limit. Below this limit the scale ambiguity suggests roughly a 10% overall normalization uncertainty.

Use of the ratio of cross sections in various rapidity bins to that in the central region further reduces this uncertainty. The dependence on the parton density function requires a more detailed study. The distribution presented in this letter may well not be the best way to study it and several relevant experimental results have already been presented by both Fermilab collaborations (see for instance ref. [16]). A next-to-leading order study of these distributions is now in order.

## Acknowledgements.

W.T.G is happy to acknowledge many stimulating conversations with the D0 collaboration and in particular with Jerry Blazey and Terry Geld. We also thank Steve Ellis for assisting with the numerical comparison with the results of ref. [8] for the single jet inclusive distribution.

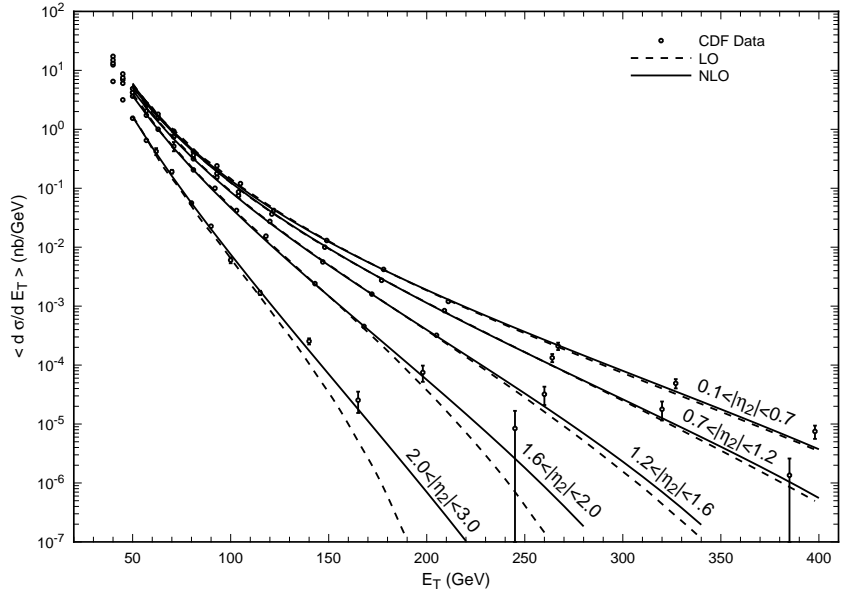

Figure 1: The leading (LO) and next-to-leading order (NLO) predictions for $\left\langle \frac{d\sigma}{dE_T} \right\rangle$ as defined in eq. 5 for $0.1 < |\eta_1| < 0.7$. The data is taken from ref. [12].

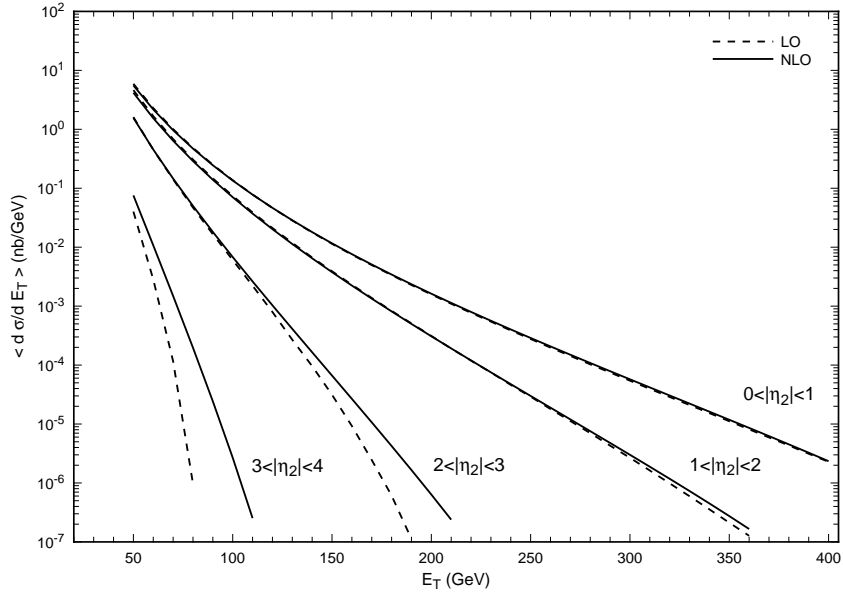

Figure 2: The leading (LO) and next-to-leading order (NLO) predictions for $\left\langle \frac{d\sigma}{dE_T} \right\rangle$ as defined in eq. 5 for $|\eta_1| < 1$.

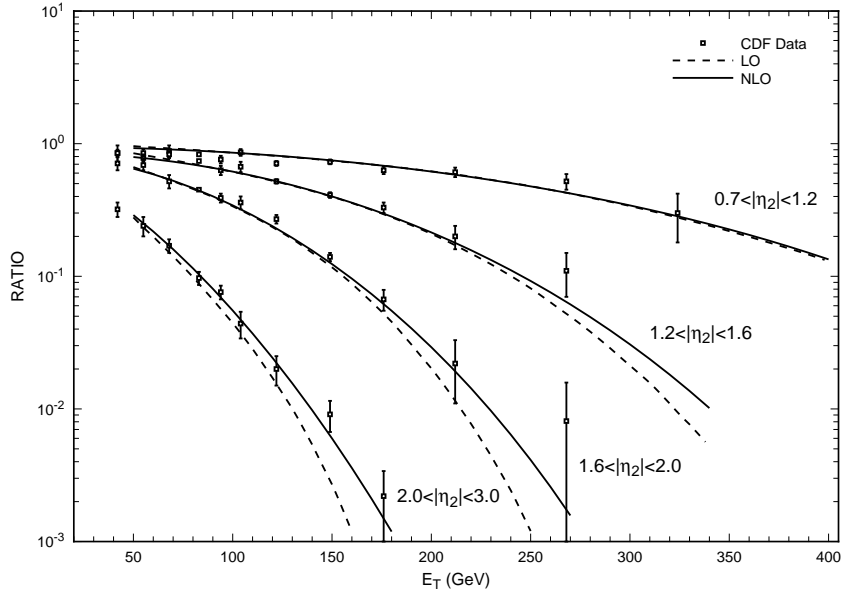

Figure 3: The leading (LO) and next-to-leading order (NLO) predictions for the ratio $R$ of the cross section for large $\eta_2$ compared to that in the central region, $0.1 < |\eta_2| < 0.7$, as defined by eq. 6 for $0.1 < |\eta_1| < 0.7$. The data is taken from ref. [12].

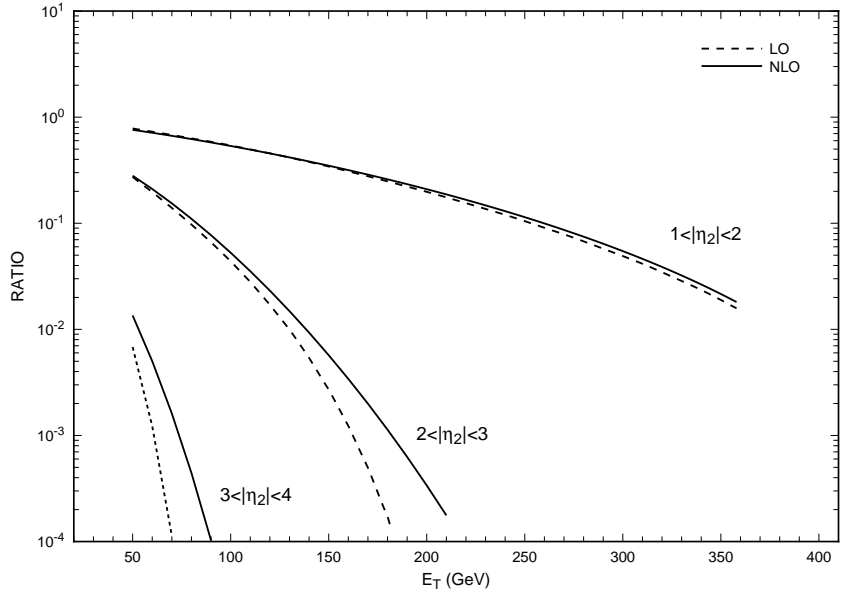

Figure 4: The leading (LO) and next-to-leading order (NLO) predictions for the ratio $R$ of the cross section for large $\eta_2$ compared to that in the central region, $|\eta_2| < 1$, as defined by eq. 6 for $|\eta_1| < 1$.

9